\documentclass[prd,amsmath,amssymb,twocolumn]{revtex4-1}

\usepackage{graphicx}
\usepackage{xcolor}
\usepackage{amsmath}
\usepackage{amssymb}

\usepackage{bm}
\usepackage{bbm}
\usepackage{MnSymbol}

\newcommand{\HO}{\mathbb{H}\otimes\mathbb{O}}
\newcommand{\RCH}{\mathbb{R}\otimes\mathbb{C}\otimes\mathbb{H}}

\newcommand{\CH}{\mathbb{C}\otimes\mathbb{H}}
\newcommand{\CO}{\mathbb{C}\otimes\mathbb{O}}
\newcommand{\CHO}{\mathbb{C}\otimes\mathbb{H}\otimes\mathbb{O}}
\newcommand{\RCHO}{\mathbb{R}\otimes\mathbb{C}\otimes\mathbb{H}\otimes\mathbb{O}}

\newcommand{\C}{\mathbb{C}}
\newcommand{\R}{\mathbb{R}}

\newcommand{\uu}{\epsilon_{\uparrow\uparrow}}
\newcommand{\ud}{\epsilon_{\uparrow\downarrow}}
\newcommand{\du}{\epsilon_{\downarrow\uparrow}}
\newcommand{\dd}{\epsilon_{\downarrow\downarrow}}

\newcommand{\gsm}{ \mathfrak{su}(3)_{\textup{C}} \oplus\mathfrak{su}(2)_{\textup{L}}\oplus\mathfrak{u}(1)_{\textup{Y}}    }

\usepackage{xcolor}

\usepackage{color}
 
\begin{document}

\title{Three Generations and a Trio of Trialities \rm}

\author{N. Furey and M.J. Hughes}
\affiliation{$ $\\  Humboldt-Universit\"{a}t zu Berlin,\\ Zum Grossen Windkanal 2, Berlin, 12489, Germany  \\ furey@physik.hu-berlin.de \\HU-EP-24/26\vspace{2mm}\\ University of Leeds, \\Woodhouse, Leeds, LS2 9JT, United Kingdom \\m.j.hughes@leeds.ac.uk  }

\begin{abstract}
We identify the Standard Model's $\mathfrak{su}(3)\oplus \mathfrak{su}(2)\oplus \mathfrak{u}(1)$ internal symmetries 
within the triality symmetries $\mathfrak{tri}(\mathbb{C}) \oplus \mathfrak{tri}(\mathbb{H}) \oplus \mathfrak{tri}(\mathbb{O})$.  From here, the corresponding Standard Model group action is  applied to the triality triple $\left( \Psi_+, \Psi_-,V\right)$ for $\Psi_+, \Psi_-, V \in \CHO$.   Together, $\Psi_+$ and $\Psi_-$ provide the correct irreducible representations for two generations.  Owing to a certain \emph{Cartan factorization,} which we define, $V$ provides  the irreducible representations for a third generation.  Said more explicitly in another way, division algebraic multiplication merges a third generation of spinor  representations into a set of scalar bosons. This set of scalar bosons includes the familiar Standard Model Higgs representation.
 \end{abstract}

\maketitle

\section{Introduction} 

Nature's three iterations of fermionic states would seem to indicate that its particle content is not merely random.  With this said, the three generation problem in particle physics is as old as the Standard Model itself, \citep{Ram1977}-\citep{Fin2024}.  Why should there be three sets of fermions that mirror a common behaviour under the Standard Model's $\gsm$ symmetries?  Perhaps one reason why the question has remained open for so long reduces to the fact that it can be difficult to find relevant mathematical objects that demand a threefold structure. 

One known threefold structure, however, goes by the name of \emph{triality}, and echos across the algebraic particle physics literature.    The idea that octonionic triality could explain the existence of three generations was volunteered in 1977 by Ramond,~\citep{Ram1977}.  Since then, a sizeable ensemble of authors have, in their own ways, reinforced the proposal.  Triality appears, for example,  in the context of the exceptional Jordan algebra, \citep{Sil1994}-\citep{Vai2023}, and in $\mathfrak{e}_8$ models, \citep{Che2020}-\citep{Lis2024}.

With this said, it is an underappreciated fact that the phenomenon of triality occurs not only for the octonions, but for all four of the finite dimensional normed division algebras over the reals.  In this article, we pursue a proposal first introduced in~\citep{FH2021(II)} to examine not only octonionic triality symmetries, $\mathfrak{tri}(\mathbb{O})$, but rather, $\mathfrak{tri}(\mathbb{R}) \oplus\mathfrak{tri}(\mathbb{C}) \oplus \mathfrak{tri}(\mathbb{H}) \oplus \mathfrak{tri}(\mathbb{O}) = \mathfrak{tri}(\mathbb{C}) \oplus \mathfrak{tri}(\mathbb{H}) \oplus \mathfrak{tri}(\mathbb{O})$.  We identify three generations of fermions with three copies of $\CHO$, labeled as 
\begin{equation}\label{3cho}
(\C_+\otimes\mathbb{H}_+\otimes\mathbb{O}_+)\oplus (\C_-\otimes\mathbb{H}_-\otimes\mathbb{O}_-) \oplus (\C_V\otimes\mathbb{H}_V\otimes\mathbb{O}_V). 
\end{equation}
\noindent For earlier work based on $\CHO$, please see \citep{Dix2004}-\citep{Jens2023}.

This article begins by explaining the  common  meanings  imparted on the term \emph{triality}.  We then identify a copy of $\gsm$ symmetries within $\mathfrak{tri}(\mathbb{H}) \oplus \mathfrak{tri}(\mathbb{O})\subset\mathfrak{tri}(\mathbb{C}) \oplus \mathfrak{tri}(\mathbb{H}) \oplus \mathfrak{tri}(\mathbb{O})$.  When applied to~(\ref{3cho}), we find that together $\C_+\otimes\mathbb{H}_+\otimes\mathbb{O}_+$ and $\C_-\otimes\mathbb{H}_-\otimes\mathbb{O}_-$ supply the correct irreducible representations for two generations, each including a sterile neutrino.  However, $\C_V\otimes\mathbb{H}_V\otimes\mathbb{O}_V$, \emph{at first sight}, does not decompose into the irreps of a third generation.  Instead, one finds a set of bosons, which oddly enough, contain the familiar Standard Model Higgs.  

How might one then unfurl a third generation?

We introduce a technique known as \emph{Cartan factorization} that allows us to recast $\C_V\otimes\mathbb{H}_V\otimes\mathbb{O}_V$ as a product of spinor and conjugate spinor representations.  Together, these spinor and conjugate spinor representations yield the states of a third generation, sterile neutrino included.  

This technique allows us to address some  challenges familiar to the problem of three generations,~\citep{Fur2023(II)}.   Namely, it allows us  to set up a three-generation triality model with the correct Standard Model charges across all three generations, whereby the three generations are seen to be linearly independent.

We then introduce \emph{Cartan factorization diagrams}, and subsequently provide a complete set of diagrams demonstrating $\C_V\otimes\mathbb{H}_V\otimes\mathbb{O}_V$'s decomposition into spinors.

This article compiles a list of possible Yukawa terms, some of which are familiar from the Standard Model, while others not.  We observe that those Yukawa terms familiar from the Standard Model comprise a special class that fulfill the requirements of a \emph{non-degenerate trilinear form}.  This leads us to propose the \emph{NDTF constraint}.

One interesting feature of $\C_V\otimes\mathbb{H}_V\otimes\mathbb{O}_V$'s original decomposition was that it included multiple copies of the Standard Model's Higgs representation.  We intend to investigate this Higgs sector, and symmetry breaking, in upcoming work.

 Finally, we mention that this model is compatible  with a Freudenthal-Tits triality construction of $\mathfrak{e}_7,$ and by extension, $\mathfrak{e}_8$,  \citep{mia}.   Explicitly, $\mathfrak{e}_7 \simeq \left(\mathfrak{tri}(\mathbb{H}) \oplus \mathfrak{tri}(\mathbb{O})\right)+ 3 \hspace{.5mm}\HO,$ and $\mathfrak{e}_8 \simeq \left(\mathfrak{tri}(\mathbb{O}) \oplus \mathfrak{tri}(\mathbb{O})\right)+ 3 \hspace{.5mm}\mathbb{O}\otimes\mathbb{O},$ with $\mathbb{H}\subset\mathbb{O}.$  To the best of our knowledge, this article provides an alternative identification of the Standard Model's three generations within $\mathfrak{e}_7,$ and $\mathfrak{e}_8,$ that has not yet appeared  in the literature.  

Our current model is also consistent with a proposed extension of the magic square based on $\mathfrak{tri}(\mathbb{C}) \oplus\mathfrak{tri}(\mathbb{H}) \oplus \mathfrak{tri}(\mathbb{O}).$  (Hodge duality and a requirement of anomaly cancellation can be implemented to reduce symmetries.) \color{black}

\section{Triality}

The term \emph{triality} is used to describe a number of closely related phenomena in algebra and in the representation theory of groups, \citep{ms}-\citep{mia}.  In this section, we introduce the reader to some common meanings of the term.  Before doing so, however, we begin by introducing the four finite dimensional normed division algebras over the reals.

\subsection{$\R,$ $\C,$ $\mathbb{H},$  $\mathbb{O}$\label{RCHOintro}}

A theorem by Hurwitz, \citep{Hur1898}, \citep{Hur1923},  states that, up to isomorphism, there exist exactly four  unital finite-dimensional normed division algebras over $\R$.  They are the real numbers, $\R$, the complex numbers, $\C$, the quaternions, $\mathbb{H},$ and the octonions, $\mathbb{O}$.

The real numbers ($\R$) are ubiquitous in physics; the complex numbers ($\C = \R\otimes\C$) are central to quantum theory; the complex quaternions, ($\CH = \RCH$), underlie Einstein's Special Relativity,~\citep{thesis}.  It is natural to then wonder about possible physical domains for $\CHO = \RCHO$.

In order to maintain notational consistency with previous work, \citep{FH2021(I)}-\citep{Fur2023(III)}, we will describe a generic complex number as $r_0 +r_1i$, where $r_0,r_1\in\R,$ $i^2 = -1$, and $(r_0+ir_1)^* = r_0-ir_1$.  

Similarly, we will describe a generic quaternion as $r_0 +r_m \epsilon_m$, for $m\in\{1,2,3\},$ where $r_0,r_m\in\R.$ Multiplicatively, $\epsilon_1\epsilon_1 = \epsilon_2\epsilon_2 = \epsilon_3\epsilon_3 = -1$.  When $m\neq n,$ we have $\epsilon_m\epsilon_n = \varepsilon_{mnp}\epsilon_p$ where $\varepsilon_{mnp}$ is the usual totally anti-symmetric tensor with $\varepsilon_{123} = 1$.  Furthermore, $\left(\widetilde{r_0+r_m\epsilon_m}\right) = r_0-r_m\epsilon_m$ with $\widetilde{ab} = \widetilde{b}\hspace{.5mm}\widetilde{a}$ for all $a,b\in \mathbb{H}$.

Finally, we will describe a generic octonion as $r_0 +r_j e_j$, for $j\in\{1,2,\dots 7\},$ where $r_0,r_j\in\R.$ Multiplicatively, $e_1e_1 = e_2e_2 = \dots = e_7e_7 = -1$. When $i\neq j,$ we have $e_i e_j = f_{ijk} e_k$, where $f_{ijk}$ is again a totally antisymmetric tensor with $f_{ijk}=1$ when $ijk\in \{124, 235, 346, 457, 561, 672, 713\}$.  Those remaining values of $f_{ijk}$ not determined by anti-symmetry are otherwise zero.  Furthermore, $\left(\widetilde{r_0+r_je_j}\right) = r_0-r_je_j$ with $\widetilde{ab} = \widetilde{b}\hspace{.5mm}\widetilde{a}$ for all $a,b\in \mathbb{O}$.  With this particular choice of indices,  octonionic multiplication enjoys an index cycling symmetry and index doubling symmetry.  Explicitly, $e_ie_{j} = e_k \Rightarrow e_{i+1}e_{j+1} = e_{k+1},$ and $e_ie_{j} = e_k \Rightarrow e_{2i}e_{2j} = e_{2k},$ respectively, where indices are understood to be elements of $\{1, 2, \dots 7\}$, more precisely, $\mathbb{Z}_7$. 

Throughout this paper, tensor products will be understood to be over $\R$, unless otherwise stated.  We will be particularly interested in algebras $\HO$ and $\CHO$.  Simplifying notation by omitting $\otimes$ symbols and trivial units, we may write a generic element of $\HO$ as $b_{0} + b_m\epsilon_m + b_j' e_j + b_{mj}'' \hspace{.5mm}\epsilon_m e_j$ for $b_0, b_m, b_j', b_{mj}''\in\R$.  Similarly, a generic element of $\CHO$ may be written again as $b_{0} + b_m\epsilon_m + b_j' e_j + b_{mj}''\hspace{.5mm} \epsilon_m e_j$,  although now with $b_0, b_m, b_j', b_{mj}''\in\C$.  Multiplication is defined on $\HO$ and $\CHO$ in the canonical way for tensor products of algebras.  As a concrete example, consider $\left(b_0 + b_1\epsilon_1 +b_{24}''\hspace{.5mm}\epsilon_2e_4\right)\left(c_0 +c_{35}''\hspace{.5mm}\epsilon_3e_5\right) = b_0c_0 +b_0c_{35}''\hspace{.5mm}\epsilon_3e_5+ b_1c_0\hspace{.5mm}\epsilon_1 - b_1c_{35}''\hspace{.5mm} \epsilon_2e_5 + b_{24}''c_0\hspace{.5mm}\epsilon_2 e_4 + b_{24}''c_{35}''\hspace{.5mm}\epsilon_1e_7$.

\subsection{Triality, as it relates to duality\label{trilin}}


We begin by introducing triality via a certain type of scalar that is formed from spinor and vector inputs familiar to physicists, 
\begin{equation}\label{triform1}t(\Psi_+, V, \Psi_- ) = \langle \Psi_+^{\dagger}\hspace{.5mm} V\hspace{.5mm} \Psi_- \rangle.
\end{equation}

Given two vector spaces $W_1$ and $W_2$ over $\mathbb{R}$, a \emph{duality} is a non-degenerate bilinear map 
\begin{equation}f:W_1\times W_2\rightarrow \mathbb{R}.  
\end{equation}
\noindent The non-degeneracy property of $f$ means that there exists no non-zero vector in one vector space such that $f$ maps to zero for all vectors in the other vector space.

Now as a natural extension, a \emph{triality} is a non-degenerate trilinear map 
\begin{equation}t :W_1\times W_2 \times W_3 \rightarrow \mathbb{R}.  
\end{equation}
\noindent The non-degeneracy property of $t$ means that there exists no two non-zero vectors in two of the vector spaces such that $t$ maps to zero for all vectors in the third vector space, \citep{baez}.

Importantly, this trilinear map may  be reformulated as a bilinear map
\begin{equation}\label{mult} m:  W_1\times W_2 \rightarrow W_3^{\star},
\end{equation}
\noindent where $W_3^{\star}$ is the dual of $W_3$.  As a result of the non-degeneracy of $t$, each of $W_1,$ $W_2$, $W_3,$ and their duals, may be identified with the same vector space $W$.  As a consequence, we see that $m$ describes multiplication,
\begin{equation}m: W\times W \rightarrow W.
\end{equation}
\noindent From here, it can be shown, \citep{baez}, that $W$ together with $m$ must be isomorphic to either $\R$, $\C$, $\mathbb{H}$, or $\mathbb{O}$.  We write $\mathbb{W}:=\left(W, m\right) \simeq \R, \C, \mathbb{H},  \mathbb{O}.$   In other words, under these conditions, \emph{triality occurs exactly four times},  one for each of the finite dimensional normed division algebras over the reals:  $\R$, $\C$, $\mathbb{H}$, and $\mathbb{O}$.

In practice, one may write down this trilinear form as 
\begin{equation}\begin{array}{ll}\label{triform}t(\Psi_+, V, \Psi_- ) &= \langle \left(\Psi_+^{\dagger}\hspace{.5mm} V\right) \Psi_- \rangle = \langle \Psi_+^{\dagger}\left( V\hspace{.5mm} \Psi_-\right) \rangle \vspace{2mm} \\
& := \langle \Psi_+^{\dagger}\hspace{.5mm} V\hspace{.5mm} \Psi_- \rangle,
\end{array}\end{equation}
\noindent for $\Psi_+\in W_1,$ $V\in W_2,$ and $\Psi_-\in W_3$.   Here, the angled brackets $\langle \cdots \rangle$ mean to take the real part.  Furthermore, $\dagger$ may be interpreted as $*$ when $\mathbb{W} = \C$, and as $ \hspace{1mm}\widetilde{   } \hspace{1mm}$ when $\mathbb{W} = \mathbb{H}$ or $\mathbb{O}.$

One then finds that for each of $\mathbb{W}=\R, \C, \mathbb{H}, \mathbb{O},$ the scalar $\langle \Psi_+^{\dagger}\hspace{.5mm} V\hspace{.5mm} \Psi_- \rangle$ is invariant under a given symmetry.  At the Lie algebra level, these symmetries are known as \emph{triality algebras}, \citep{baez}, \citep{mia}, and are given by
\begin{equation}\begin{array}{l}\label{ta}
\mathfrak{tri}(\R) \simeq \emptyset, \vspace{2mm}\\
\mathfrak{tri}(\C) \simeq \mathfrak{u}(1)\oplus\mathfrak{u}(1),\vspace{2mm}\\
\mathfrak{tri}(\mathbb{H}) \simeq \mathfrak{su}(2)\oplus\mathfrak{su}(2)\oplus\mathfrak{su}(2),\vspace{2mm}\\
\mathfrak{tri}(\mathbb{O}) \simeq \mathfrak{so}(8).
\end{array}\end{equation}
\noindent  In the next subsection, we describe the specific transformation properties of $\Psi_+,$ $V,$ and $\Psi_-$ under these symmetries.

It may be noted that $\mathfrak{tri}(\mathbb{W})$ contains the norm-preserving symmetries $\mathfrak{so}(1)\simeq \emptyset,$  $\mathfrak{so}(2) \simeq \mathfrak{u}(1),$ $\mathfrak{so}(4) \simeq \mathfrak{su}(2)\oplus\mathfrak{su}(2),$ and $\mathfrak{so}(8)$ for $\mathbb{W}=\R, \C, \mathbb{H}, \mathbb{O},$ respectively.


\subsubsection{Hughes' Higgs}

Recently, a novel physical interpretation was found by  Hughes, \citep{FH2021(II)}, for this triality scalar in the case of $\mathbb{W}=\mathbb{H}$.  Namely, she identified 
$\mathfrak{tri}(\mathbb{H})$ with $\mathfrak{su}(2)_{\textup{L}}\oplus\mathfrak{su}(2)_{\textup{R}}\oplus\mathfrak{su}(2)_{\textup{spin}}$,  left-handed fermions with $\Psi_- = \Psi_{\textup{L}}\simeq \left( \bf{2}, \bf{1}, \bf{2} \right)$, right-handed fermions as $\Psi_+ = \Psi_{\textup{R}}\simeq \left( \bf{1}, \bf{2}, \bf{2} \right)$, and a left-right symmetric Higgs  as $V = \Phi \simeq \left( \bf{2}, \bf{2}, \bf{1} \right)$.  The triality scalar, $\langle \Psi_+^{\dagger}\hspace{.5mm} V\hspace{.5mm} \Psi_- \rangle$ then embodies a Yukawa coupling.  Upon breaking $\mathfrak{su}(2)_{\textup{R}}$,  Hughes' left-right symmetric Higgs reduces to the familiar Standard Model Higgs.

\subsection{Decomposition via derivations}

Although we have listed the triality Lie algebras under which $\langle \Psi_+^{\dagger}\hspace{.5mm} V\hspace{.5mm} \Psi_- \rangle$ is invariant, we still have yet to specify exactly how $\Psi_+,$ $\Psi_-,$ and $V$ transform under these symmetries in the general case.  In order to write down these transformation rules, it will be helpful to make use of a certain decomposition of $\mathfrak{tri}(\mathbb{W})$.

The triality algebras listed in equations~(\ref{ta}) follow a certain pattern.  Namely, for $\mathbb{W} = \R$, $\C$, $\mathbb{H}$, or $\mathbb{O}$, one finds that
\begin{equation} \label{im}\mathfrak{tri}(\mathbb{W}) = \mathfrak{der}(\mathbb{W}) + \mathfrak{Im}(\mathbb{W}) + \mathfrak{Im}(\mathbb{W}).
\end{equation}
\noindent The term $\mathfrak{Im}(\mathbb{W})$ refers to the imaginary part of $\mathbb{W}$, with generic elements $r_1i\in\mathfrak{Im}(\mathbb{C})\simeq \mathfrak{u}(1)$,  $r_m\epsilon_m\in\mathfrak{Im}(\mathbb{H})\simeq \mathfrak{su}(2)$, and $r_je_j\in\mathfrak{Im}(\mathbb{O})$.  The term $\mathfrak{der}(\mathbb{W})$ refers to the \emph{derivation algebra} of $\mathbb{W}$.  

We define the \emph{endomorphisms} of an algebra $\mathbb{A}$, denoted $End(\mathbb{A})$, to be the set of all (not necessarily invertible) linear maps from $\mathbb{A}\rightarrow \mathbb{A}$.  Then $\widehat{d}\in End(\mathbb{A})$ is a \emph{derivation} if
\begin{equation}
\widehat{d}(a_1a_2) = \widehat{d}(a_1)\hspace{.5mm}a_2 + a_1\hspace{.5mm}\widehat{d}(a_2)\hspace{1cm} \forall a_1, a_2 \in\mathbb{A}.
\end{equation}
\noindent In the cases of $\mathbb{A} = \R, \C, \mathbb{H}, \mathbb{O},$ we have
\begin{equation}
\begin{array}{l}\label{der}
\mathfrak{der}(\R) \simeq \emptyset, \vspace{2mm}\\
\mathfrak{der}(\C) \simeq \emptyset,\vspace{2mm}\\
\mathfrak{der}(\mathbb{H}) \simeq \mathfrak{su}(2),\vspace{2mm}\\
\mathfrak{der}(\mathbb{O}) \simeq \mathfrak{g}_2.
\end{array}
\end{equation}
\noindent Explicit descriptions of $\mathfrak{der}(\mathbb{H}) $ and $\mathfrak{der}(\mathbb{O}) $ actions may be found in equations~(\ref{derH}) and (\ref{derO}) later on in this manuscript.

Finally, \citep{mia}, we may now set $V,$ $\Psi_+,$ and $\Psi_-^{\dagger}$ to transform  as the vector, spinor, and conjugate spinor representations for the triality symmetries of~(\ref{im}):
\begin{equation} \label{trans}
\begin{array}{lll}
\delta_V V  &=& \widehat{d}\hspace{.9mm}V+L_a \hspace{.5mm}V+ R_b\hspace{.5mm}V ,\hspace{.5mm} \vspace{2mm}\\
\delta_+ \Psi_{+} &=& \widehat{d}\hspace{.9mm}\Psi_{+} +L_a \hspace{.5mm}\Psi_{+} + R_{a-b}\hspace{.5mm}\Psi_{+},\vspace{2.2mm}\\
\delta_- {\Psi}_{-}^{\dagger}  &=& \widehat{d}\hspace{.9mm}{\Psi}_{-}^{\dagger}+L_{b-a}\hspace{.5mm}{\Psi}_{-}^{\dagger} + R_b\hspace{.5mm}{\Psi}_{-}^{\dagger},
\end{array}\end{equation}
\noindent   in which case, the invariance of $\langle \Psi_+^{\dagger}\hspace{.5mm} V\hspace{.5mm} \Psi_- \rangle$ under triality symmetries~(\ref{ta}) may be confirmed.  Here,  $a,b \in \mathfrak{Im}(\mathbb{W})$, and $L_{w_1} w_2:=w_1w_2$, while $R_{w_1} w_2:=w_2w_1$ for all $w_1,w_2\in\mathbb{W}$.  

It is straightforward to demonstrate that $\mathfrak{der}(\mathbb{W})$ constitutes the Lie subalgebra of $\mathfrak{tri}(\mathbb{W})$ whereby the spinor, conjugate spinor, and vector representations coincide.  For an interesting occurrence of a somewhat reminiscent phenomenon, see \citep{Fur2023(III)}.

\subsection{Cartan's Triality Principle \label{CTPsection}}

Thus far, we have discussed those transformations on $ \{\Psi_{+},  V, \Psi_{-} \}$ that hold $t( \Psi_{+},  V, \Psi_{-})$ fixed.  However, as mentioned earlier in equation  (\ref{mult}), this trilinear map, $t$, may be reformulated as a bilinear map, $m$.  It is in this new context that \emph{Cartan's Triality Principle} applies.

Let us suppose that $V=\Psi_+ \Psi_-^{\dagger}$, where $\Psi_+$ and  $\Psi_-^{\dagger}$ are multiplied via division algebraic multiplication, $m.$  Suppose that $\Psi_+$ and $\Psi_-^{\dagger}$ transform according to equations~(\ref{trans}).  \emph{Cartan's Triality Principle} (CTP) then states that 
\begin{equation} \label{CTP1}  \delta_V V =  \left( \delta_+ \Psi_+\right) \Psi_-^{\dagger} + \Psi_+  \left(\delta_-\Psi_-^{\dagger}\right).
\end{equation}
\noindent In other words,  the vector representation results from $m$-multiplying the spinor and conjugate spinor representations, \citep{lounesto}. 

Alternatively, suppose that $\Psi_+ = V\hspace{.5mm} \Psi_-$, with $V$ and $\Psi_-$ transforming according to equations~(\ref{trans}).  Then CTP states that 
\begin{equation} \label{CTP2}  \delta_+ \Psi_+ =  \left( \delta_V V\right) \Psi_- + V  \left(\delta_-\Psi_-^{\dagger}\right)^{\dagger}.
\end{equation}
\noindent In other words,  the spinor representation results from $m$-multiplying the vector and conjugate spinor representations. 

 Finally, suppose that $\Psi_-^{\dagger} =\Psi_+^{\dagger}\hspace{.5mm} V$ with $\Psi_+^{\dagger}$ and $V$ transforming according to equations~(\ref{trans}). Then CTP states that 
\begin{equation} \label{CTP3}  \delta_- \Psi_-^{\dagger} =  \left( \delta_+ \Psi_+\right)^{\dagger} V + \Psi_+^{\dagger}  \left(\delta_VV\right).
\end{equation}
\noindent In other words,  the conjugate spinor representation results from $m$-multiplying the spinor and vector representations.

 These division algebraic factorizations of triality representations 
\begin{equation}\begin{array}{lcl} \label{factors}
V=\Psi_+ \Psi_-^{\dagger} &\hspace{2mm}\Leftrightarrow\hspace{2mm}& V^{\dagger}=\Psi_- \Psi_+^{\dagger}, \vspace{2mm}\\
\Psi_+ = V\hspace{.5mm} \Psi_- &\Leftrightarrow&   \Psi_+^{\dagger} =  \Psi_-^{\dagger} \hspace{.5mm}V^{\dagger},\vspace{2mm}\\
\Psi_-^{\dagger} =\Psi_+^{\dagger}\hspace{.5mm}V &\Leftrightarrow & \Psi_- =V^{\dagger}\hspace{.5mm}\Psi_+
\end{array}\end{equation}
\noindent will be of special importance; we will refer to them as \emph{Cartan factorizations}.  In particular, the first of these three factorizations, the \emph{Cartan vector factorization}, will allow us to identify a third generation of Standard Model spinors from within a set of bosons that includes Higgs representations.

\subsection{Triality as the $S_3$ permutations of $\Psi_+,$ $\Psi_-,$ $V$ }

Apart from the non-degenerate trilinear form described in Subsection~(\ref{trilin}), and apart from CTP, there is yet another meaning for the term \emph{triality}.  \it Triality \rm is also commonly used to refer to a certain $S_3$ permutation symmetry of the triality representations acting on $\Psi_{+}$, $\Psi_{-}$, and $V.$

Reorganizing for convenience the operators and daggers of equations~(\ref{trans}), let us define the triality algebra actions on $V, \Psi_{-}, {\Psi}_{+}^{\dagger}$  as
\begin{equation} \begin{array}{rcl}
\Delta_V  &:=& \widehat{d}\hspace{.5mm}+L_a + R_b, \vspace{2mm}\\
\Delta_{-} &:=& \widehat{d} \hspace{.5mm}-L_b  + R_{a-b},\vspace{2mm}\\
\Delta_{+}  &:=& \widehat{d}\hspace{.5mm}+L_{b-a} - R_a,
\end{array}
\end{equation}
\noindent respectively.  From here, one finds that repeated iterations of the map 
\begin{equation} \begin{array}{llll}\gamma : \hspace{2mm}&a&\mapsto &-b, \vspace{1mm}\\
&b&\mapsto &a-b.
\end{array}\end{equation}
\noindent cycles $\Delta_V \mapsto \Delta_{-}\mapsto \Delta_{+}\mapsto \Delta_{V}.$  This single map, $\gamma,$ generates the three-element cyclic group $C_3\simeq \mathbb{Z}_3 \subset S_3$.

\begin{figure}[h!]
\begin{center}
\includegraphics[width=8.5cm]{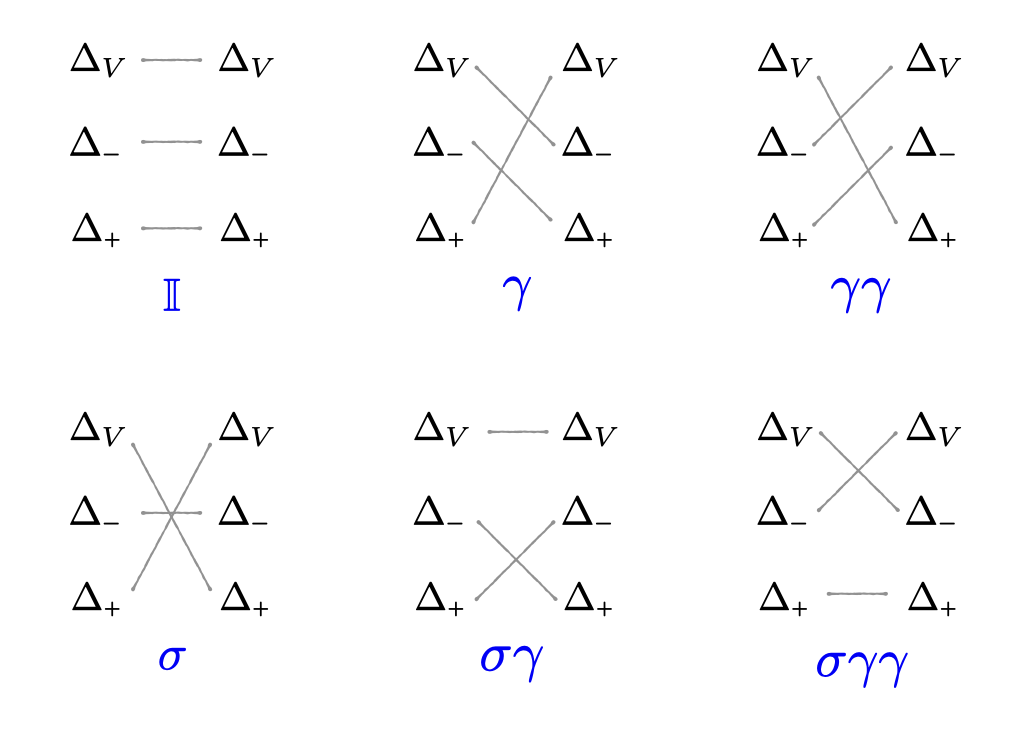}
\caption{\label{S3}
Triality's $S_3$ permutations between vector, $\Delta_V$, conjugate spinor, $\Delta_-$, and spinor $\Delta_+$ actions.}
\end{center}
\end{figure}

In order to generate the full $S_3$ group, we introduce as well an adjacent transposition map, $\sigma$, defined as
\begin{equation}\begin{array}{llll}\sigma : \hspace{2mm}&a&\mapsto &a, \vspace{1mm}\\
&b&\mapsto &a-b.
\end{array}
\end{equation}
\noindent This map swaps spinor and vector  representations.  Careful readers will notice that $\sigma$ is indeed an adjacent transposition map up to an overall application of the anti-involution map, $\dagger$, as per equations~(\ref{factors}).  Figure~\ref{S3} demonstrates the $S_3$ triality action on $\Delta_V,$ $\Delta_-, $ and $\Delta_+$ representations.

\section{Multiple Simultaneous Trialities}


In this section we will  first consider the tensor product $\HO$.  This set up will lead us towards Standard Model fermions in a Majorana representation, as opposed to the usual Weyl representation.  The fact that such a Majorana description is possible (not only for sterile neutrinos) has been demonstrated extensively in~\citep{BM}.  Readers may also find reference~\citep{BBlog} and \citep{thesis} helpful.  

Let us identify $\mathfrak{su}(2)_{\textup{L}}$-active Majorana fermions with $\mathbb{H}_-\otimes\hspace{.5mm}\mathbb{O}_-$ and $\mathfrak{su}(2)_{\textup{L}}$-inactive Majorana fermions with $\mathbb{H}_+\otimes\hspace{.5mm}\mathbb{O}_+$.  We will then show three generations of Standard Model fermion irreps as ultimately originating from $$(\mathbb{H}_+\otimes\mathbb{O}_+)\oplus (\mathbb{H}_-\otimes\mathbb{O}_-) \oplus (\mathbb{H}_V\otimes\mathbb{O}_V).$$

\subsection{$\mathfrak{tri}(\mathbb{O})\oplus \mathfrak{tri}(\mathbb{H})$ actions}

Suppose that $\mathcal{H}_+\in\mathbb{H}_+,$  $\mathcal{H}_-\in\mathbb{H}_-,$ $\mathcal{H}_V\in\mathbb{H}_V,$ $\mathcal{O}_+\in\mathbb{O}_+,$  $\mathcal{O}_-\in\mathbb{O}_-,$ $\mathcal{O}_V\in\mathbb{O}_V,$ where the subscripts allow us to distinguish spinor, conjugate spinor, and vector representations.  
 Then a generic element of $\mathbb{H}_+\otimes\mathbb{O}_+$ may be written a sum of objects of the form $\mathcal{H}_+\otimes\mathcal{O}_+$.  Analogous statements hold true for $\mathbb{H}_-\otimes\mathbb{O}_-$ and $\mathbb{H}_V\otimes\mathbb{O}_V$.  
Omitting $\otimes$ symbols between $\mathcal{H}$ and $\mathcal{O}$, for notational simplicity,  these objects transform as
\begin{equation}\begin{array}{lcl}
\delta_V \left(\mathcal{H}_V\mathcal{O}_V\right) &=& \delta^V_{\mathbb{H}}\left( \mathcal{H}_V\right)\mathcal{O}_V + \mathcal{H}_V\hspace{.5mm}\delta^V_{\mathbb{O}}\left( \mathcal{O}_V\right)  \vspace{2.2mm}\\
\delta_+ \left(\mathcal{H}_+\mathcal{O}_+\right) &=& \delta^+_{\mathbb{H}}\left( \mathcal{H}_+\right)\mathcal{O}_+ + \mathcal{H}_+\hspace{.5mm}\delta^+_{\mathbb{O}}\left( \mathcal{O}_+\right) \vspace{2mm}\\
\delta_- \left(\mathcal{H}_-\mathcal{O}_-\right)^{\dagger} &=& \delta^-_{\mathbb{H}}\left( \mathcal{H}_-^{\dagger}\right)\mathcal{O}_-^{\dagger} + \mathcal{H}_-^{\dagger}\hspace{.5mm}\delta^-_{\mathbb{O}}\left( \mathcal{O}_-^{\dagger}\right),
\end{array}\end{equation}
\noindent where 
\begin{equation} \label{htrans}
\begin{array}{lll}
\delta^V_{\mathbb{H}}   &=& \widehat{d}_{\mathbb{H}}+L_a + R_b\ , \vspace{2mm}\\
\delta^+_{\mathbb{H}}  &=& \widehat{d}_{\mathbb{H}} +L_a  + R_{a-b},\vspace{2.2mm}\\
\delta^-_{\mathbb{H}}   &=& \widehat{d}_{\mathbb{H}}+L_{b-a} + R_b
\end{array}\end{equation}
\noindent for $a,b\in\mathfrak{Im}(\mathbb{H})$, and 
\begin{equation} \label{otrans}
\begin{array}{lll}
\delta^V_{\mathbb{O}}   &=& \widehat{d}_{\mathbb{O}}+L_{\alpha} + R_{\beta}\ , \vspace{2mm}\\
\delta^+_{\mathbb{O}}  &=& \widehat{d}_{\mathbb{O}} +L_{\alpha}  + R_{\alpha - \beta},\vspace{2.2mm}\\
\delta^-_{\mathbb{O}}   &=& \widehat{d}_{\mathbb{O}}+L_{\beta-\alpha} + R_{\beta}
\end{array}\end{equation}
\noindent for $\alpha, \beta \in\mathfrak{Im}(\mathbb{O}).$  

The quaternionic derivation algebra $\mathfrak{der}(\mathbb{H}) = \mathfrak{su}(2)$ acts on $\mathbb{H}$ as 
\begin{equation} \label{derH} \widehat{d}_{\mathbb{H}}:= L_{r}-R_{r}
\end{equation}
\noindent for $r\in\mathfrak{Im}(\mathbb{H})$, while the octonionic derivation algebra $\mathfrak{der}(\mathbb{O}) =\mathfrak{g}_2$ acts on $\mathbb{O}$ as

\begin{equation}\begin{array}{lll}\label{derO}
 \widehat{d}_{\mathbb{O}}&:=&  \frac{\rho_1}{2}\left(L_{34}-L_{15} \right) + \frac{\rho_2}{2}\left(L_{14}+L_{35} \right) \vspace{2mm} \\
 &+& \frac{\rho_3}{2}\left(L_{13}-L_{45} \right) -\frac{\rho_4}{2}\left(L_{25}+L_{46} \right) \vspace{2mm} \\
 &+& \frac{\rho_5}{2}\left(L_{24}-L_{56} \right) -\frac{\rho_6}{2}\left(L_{16}+L_{23} \right)\vspace{2mm} \\
 &-&  \frac{\rho_7}{2}\left(L_{12}+L_{36} \right) - \frac{\rho_8}{2\sqrt{3}}\left(L_{13}+L_{45}  -  2L_{26} \right)\vspace{2mm} \\
&+&  \frac{\rho_9}{2\sqrt{3}}\left( L_{15}+L_{34}+2L_{27}\right) + \frac{\rho_{10}}{2\sqrt{3}}\left( -L_{14}+L_{35}-2L_{67}\right)\vspace{2mm}\\
&+&  \frac{\rho_{11}}{2\sqrt{3}}\left( L_{46}-L_{25}+2L_{17}\right) +\frac{\rho_{12}}{2\sqrt{3}}\left( L_{24}+L_{56}-2L_{37}\right)\vspace{2mm}\\
&+& \frac{ \rho_{13}}{2\sqrt{3}}\left( -L_{16}+L_{23}+2L_{47}\right) + \frac{\rho_{14}}{2\sqrt{3}}\left( L_{12}-L_{36}-2L_{57}\right),\vspace{1mm}
\end{array}\end{equation}
\noindent where $\rho_k \in \R$ and $L_{ij}$ is shorthand for $L_{e_i}L_{e_j}$.

\subsection{Standard Model symmetries inside $\mathfrak{tri}(\mathbb{O})\oplus \mathfrak{tri}(\mathbb{H})$}

With the actions of $\mathfrak{tri}(\mathbb{O})$ and $\mathfrak{tri}(\mathbb{H})$  defined, we may now identify a copy of $\mathfrak{su}(3)_{\textup{C}}\oplus \mathfrak{su}(2)_{\textup{L}}\oplus \mathfrak{u}(1)_{\textup{Y}}$ within them.

It has been known since at least the 1970s that $\mathfrak{su}(3)_{\textup{C}}$ may be identified within $\mathfrak{g}_2$ as the Lie subalgebra fixing an octonionic imaginary unit,~\citep{GGquarks}.  Readers may confirm that setting $\rho_9$  to $\rho_{14}$ of equation~(\ref{derO}) to zero defines an $\mathfrak{su}(3)_{\textup{C}}$ subalgebra that fixes the octonionic imaginary unit $e_7$.  We are then left with
\begin{equation}\begin{array}{lll}
\delta_{\mathfrak{su}(3)_{\textup{C}}}^{+} &=& \delta_{\mathfrak{su}(3)_{\textup{C}}}^{-} \hspace{1mm}=\hspace{1mm}\delta_{\mathfrak{su}(3)_{\textup{C}}}^{V}\vspace{2mm} \\
&=&  \frac{\rho_1}{2}\left(L_{34}-L_{15} \right) + \frac{\rho_2}{2}\left(L_{14}+L_{35} \right) \vspace{2mm} \\
 &+& \frac{\rho_3}{2}\left(L_{13}-L_{45} \right) -\frac{\rho_4}{2}\left(L_{25}+L_{46} \right) \vspace{2mm} \\
 &+& \frac{\rho_5}{2}\left(L_{24}-L_{56} \right) -\frac{\rho_6}{2}\left(L_{16}+L_{23} \right)\vspace{2mm} \\
 &-&  \frac{\rho_7}{2}\left(L_{12}+L_{36} \right) - \frac{\rho_8}{2\sqrt{3}}\left(L_{13}+L_{45}  -  2L_{26} \right).
\end{array}\end{equation}
\noindent  Given that $\mathfrak{su}(3)_{\textup{C}}\subset \mathfrak{der}({\mathbb{O}})$, one finds that $\mathfrak{su}(3)_{\textup{C}}$ acts identically on the spinor, conjugate spinor, and vector representations.

On the other hand, we will identify $\mathfrak{su}(2)_{\textup{L}}$ with the $\mathfrak{tri}(\mathbb{H})$ Lie subalgebra for which $a=-r$ and $b=-2r$.  For spinor, conjugate spinor, and vector representations,  $\mathfrak{su}(2)_{\textup{L}}$ acts as
\begin{equation}
\delta_{\mathfrak{su}(2)_{\textup{L}}}^{+} = 0, \hspace{5mm} \delta_{\mathfrak{su}(2)_{\textup{L}}}^{-} = -R_{3r}, \hspace{5mm} \delta_{\mathfrak{su}(2)_{\textup{L}}}^{V} = -R_{3r},
\end{equation}
\noindent with two of the three representations transforming identically.

Using the weak hypercharge conventions of~\citep{BM}, we may define $\mathfrak{u}(1)_Y$ within $\mathfrak{tri}(\mathbb{O})\oplus \mathfrak{tri}(\mathbb{H})$ by setting $\rho_i =0$ $\forall i$, $\alpha = \frac{y}{6}e_7$, $\beta = -\frac{y}{6}e_7$, $r=b=-\frac{y}{6}\epsilon_3$, and $a=-\frac{y}{3}\epsilon_3$ for $y\in\R$.  This results in the weak hypercharge actions
\begin{equation} \label{Y}
\begin{array}{lll}
\delta^+_{\mathfrak{u}(1)_Y}  &=& L_{\frac{y}{6}e_7-\frac{y}{2}\epsilon_3}  +R_{\frac{y}{3}e_7},\vspace{2.2mm}\\
\delta^-_{\mathfrak{u}(1)_Y}   &=& L_{-\frac{y}{3}e_7}  +R_{-\frac{y}{6}e_7},\vspace{2.2mm}\\
\delta^V_{\mathfrak{u}(1)_Y}   &=& L_{\frac{y}{6}e_7-\frac{y}{2}\epsilon_3}  +R_{-\frac{y}{6}e_7},
\end{array}\end{equation}
\noindent which behave differently on each representation.  It is also worth noting that weak hypercharge here is described using  a single octonionic, and a single quaternionic imaginary unit.

\subsection{$B-L$ and electric charge $Q$}

Likewise, $\mathfrak{u}(1)_{B-L}$ and $\mathfrak{u}(1)_{Q}$ may be found within $\mathfrak{tri}(\mathbb{O})\oplus \mathfrak{tri}(\mathbb{H})$, and act as
\begin{equation} \label{B-L}
\begin{array}{lll}
\delta^+_{\mathfrak{u}(1)_{B-L}}  &=& L_{\frac{\ell}{3}e_7}  +R_{\frac{2\ell}{3}e_7},\vspace{2.2mm}\\
\delta^-_{\mathfrak{u}(1)_{B-L}}   &=& L_{-\frac{2\ell}{3}e_7}  +R_{-\frac{\ell}{3}e_7},\vspace{2.2mm}\\
\color{black}\delta^V_{\mathfrak{u}(1)_{B-L}}   &\color{black}=& \color{black}L_{\frac{\ell}{3}e_7}  +R_{-\frac{\ell}{3}e_7},\color{black}
\end{array}\end{equation}
\noindent and
\begin{equation} \label{Q}
\begin{array}{lll}
\delta^+_{\mathfrak{u}(1)_{Q}}  &=& L_{\frac{q}{6}e_7-\frac{q}{2}\epsilon_3}  +R_{\frac{q}{3}e_7},\vspace{2.2mm}\\
\delta^-_{\mathfrak{u}(1)_{Q}}   &=& L_{-\frac{q}{3}e_7}  +R_{-\frac{q}{6}e_7+\frac{q}{2}\epsilon_3},\vspace{2.2mm}\\
\color{black}\delta^V_{\mathfrak{u}(1)_{Q}}   &\color{black}=&\color{black} L_{\frac{q}{6}e_7-\frac{q}{2}\epsilon_3}  +R_{-\frac{q}{6}e_7+ \frac{q}{2}\epsilon_3},\color{black}
\end{array}\end{equation}
\noindent for $\ell, q \in\R$.  The vector representations for these symmetries are worth pointing out.  That is, $\delta^V_{\mathfrak{u}(1)_{B-L}}$ is described simply by a commutator with $e_7$, whereas $\delta^V_{\mathfrak{u}(1)_{Q}}$ is described by a commutator with a certain linear combination of $e_7$ and $\epsilon_3$.   


\subsection{First two generations}

With spinor, conjugate spinor, and vector representations of $\gsm$ defined, we may now begin identifying fermion states within 
$$(\mathbb{H}_+\otimes\mathbb{O}_+)\oplus (\mathbb{H}_-\otimes\mathbb{O}_-) \oplus (\mathbb{H}_V\otimes\mathbb{O}_V).$$

In keeping with previous articles,  \citep{thesis}, \citep{malala}, \citep{FH2021(I)}, \citep{FH2021(II)}, \citep{Fur2023(I)},  we will define $\CO$ basis vectors as
\begin{equation}\begin{array}{ll} 
{\ell}:=\frac{1}{2}\left(1+ie_7\right),\hspace{4mm}&{\ell}^*:=\frac{1}{2}\left(1-ie_7\right),\vspace{2mm}\\
\alpha_1:= \frac{1}{2}\left(-e_5+ie_4\right),\hspace{4mm} &\alpha_1^{\dagger}:= \frac{1}{2}\left(e_5+ie_4\right),\vspace{2mm}\\
\alpha_2:= \frac{1}{2}\left(-e_3+ie_1\right),\hspace{4mm} &\alpha_2^{\dagger}:= \frac{1}{2}\left(e_3+ie_1\right),\vspace{2mm}\\
\alpha_3:= \frac{1}{2}\left(-e_6+ie_2\right),\hspace{4mm} &\alpha_3^{\dagger}:= \frac{1}{2}\left(e_6+ie_2\right),
\end{array}\end{equation}
\noindent and $\CH$ basis vectors as
\begin{equation}\begin{array}{ll} 
\uu:=\frac{1}{2}\left(1+i\epsilon_3\right), \hspace{4mm}& \ud:=\frac{1}{2}\left(-\epsilon_2+i\epsilon_1\right),\vspace{2mm}\\
\du:=\frac{1}{2}\left(\epsilon_2+i\epsilon_1\right), \hspace{4mm}& \dd:=\frac{1}{2}\left(1-i\epsilon_3\right).
\end{array}\end{equation}
\noindent It may be noticed that we are temporarily introducing the complex $i\in\C$, however rest assured that $i$ drops out in the result.  

Two generations of $\mathfrak{su}(2)_{\textup{L}}$-inactive states may be identified within $\mathbb{H}_+\otimes\mathbb{O}_+$ as
\begin{equation}\begin{array}{lll} \label{++}
\mathbb{H}_+\otimes\mathbb{O}_+ &=&  \mathcal{V}^1_{\textup{R}}\hspace{.7mm} {\ell}\uu + \mathcal{V}^{1*}_{\textup{R}} \hspace{.7mm}{\ell}^*\dd \vspace{2mm}\\
&+&  \mathcal{V}^2_{\textup{R}}\hspace{.7mm} {\ell}\ud - \mathcal{V}^{2*}_{\textup{R}} \hspace{.7mm}{\ell}^*\du \vspace{2mm}\\
&+&  \mathcal{E}^{1}_{\textup{R}}\hspace{.7mm} {\ell}\du - \mathcal{E}^{1*}_{\textup{R}} \hspace{.7mm}{\ell}^*\ud \vspace{2mm}\\
&+&  \mathcal{E}^{2}_{\textup{R}}\hspace{.7mm} {\ell}\dd + \mathcal{E}^{2*}_{\textup{R}} \hspace{.7mm}{\ell}^*\uu \vspace{2mm}\\
&+&  \mathcal{U}^{i1}_{\textup{R}}\hspace{.7mm} {\alpha_i}\uu - \mathcal{U}^{i1*}_{\textup{R}} \hspace{.7mm}\alpha_i^{\dagger}\dd \vspace{2mm}\\
&+&  \mathcal{U}^{i2}_{\textup{R}}\hspace{.7mm} {\alpha_i}\ud + \mathcal{U}^{i2*}_{\textup{R}} \hspace{.7mm}\alpha_i^{\dagger}\du \vspace{2mm}\\
&+&  \mathcal{D}^{i1}_{\textup{R}}\hspace{.7mm} {\alpha_i}\du + \mathcal{D}^{i1*}_{\textup{R}} \hspace{.7mm}\alpha_i^{\dagger}\ud \vspace{2mm}\\
&+&  \mathcal{D}^{i2}_{\textup{R}}\hspace{.7mm} {\alpha_i}\dd - \mathcal{D}^{i2*}_{\textup{R}} \hspace{.7mm}\alpha_i^{\dagger}\uu,\vspace{2mm}\\

\end{array}\end{equation}
\noindent where $\mathcal{V}^1_{\textup{R}}, \mathcal{V}^2_{\textup{R}},  \mathcal{E}^{1}_{\textup{R}}, \dots \mathcal{D}^{i2}_{\textup{R}}$ are complex coefficients.  At this stage, readers may already confirm that all factors of the complex $i$ drop out. In the labeling of fermions throughout this text, we have opted against identifying whether particles come from the first, second, or third generation.  This will only be appropriate once masses are assigned.   


In an analogous fashion, we find that two generations of $\mathfrak{su}(2)_{\textup{L}}$-active states may be identified within $\mathbb{H}_-\otimes\mathbb{O}_-$ as
\begin{equation}\begin{array}{lll} \label{--}
\mathbb{H}_-\otimes\mathbb{O}_- &=&  \mathcal{V}^1_{\textup{L}}\hspace{.7mm} {\ell}\uu + \mathcal{V}^{1*}_{\textup{L}} \hspace{.7mm}{\ell}^*\dd \vspace{2mm}\\
&+&  \mathcal{V}^2_{\textup{L}}\hspace{.7mm} {\ell}\ud - \mathcal{V}^{2*}_{\textup{L}} \hspace{.7mm}{\ell}^*\du \vspace{2mm}\\
&+&  \mathcal{E}^{1}_{\textup{L}}\hspace{.7mm} {\ell}\du - \mathcal{E}^{1*}_{\textup{L}} \hspace{.7mm}{\ell}^*\ud \vspace{2mm}\\
&+&  \mathcal{E}^{2}_{\textup{L}}\hspace{.7mm} {\ell}\dd + \mathcal{E}^{2*}_{\textup{L}} \hspace{.7mm}{\ell}^*\uu \vspace{2mm}\\
&+&  \mathcal{U}^{i1}_{\textup{L}}\hspace{.7mm} {\alpha_i}\uu - \mathcal{U}^{i1*}_{\textup{L}} \hspace{.7mm}\alpha_i^{\dagger}\dd \vspace{2mm}\\
&+&  \mathcal{U}^{i2}_{\textup{L}}\hspace{.7mm} {\alpha_i}\ud + \mathcal{U}^{i2*}_{\textup{L}} \hspace{.7mm}\alpha_i^{\dagger}\du \vspace{2mm}\\
&+&  \mathcal{D}^{i1}_{\textup{L}}\hspace{.7mm} {\alpha_i}\du + \mathcal{D}^{i1*}_{\textup{L}} \hspace{.7mm}\alpha_i^{\dagger}\ud \vspace{2mm}\\
&+&  \mathcal{D}^{i2}_{\textup{L}}\hspace{.7mm} {\alpha_i}\dd - \mathcal{D}^{i2*}_{\textup{L}} \hspace{.7mm}\alpha_i^{\dagger}\uu,\vspace{2mm}\\
\end{array}\end{equation}
\noindent where $\mathcal{V}^1_{\textup{L}}, \mathcal{V}^2_{\textup{L}},  \mathcal{E}^{1}_{\textup{L}}, \dots \mathcal{D}^{i2}_{\textup{L}} \in \C.$

\subsection{An obscured third generation}

Between $\mathbb{H}_+\otimes\mathbb{O}_+$ and $\mathbb{H}_-\otimes\mathbb{O}_-$, we have now accounted for two generations.  Clearly, we would like $\mathbb{H}_V\otimes\mathbb{O}_V$ to account for a third generation.  However, calculation of  electroweak transformations, $\delta_{\mathfrak{su}(2)_{\textup{L}}}^{V}$ and $\delta^V_{\mathfrak{u}(1)_Y}$ results in what would look, na\"{i}vely,  to be incorrect charges.  That is, we  obtain instead the real representations corresponding to the familiar $( \mathfrak{su}(3)_{\textup{C}},  \mathfrak{su}(2)_{\textup{L}}, \mathfrak{u}(1)_{\textup{Y}} )$ complex representations 
\begin{equation}\label{scalars}
\left( \mathbf{1}, \mathbf{2}, \frac{1}{2} \right), \hspace{2mm}\left( \mathbf{1}, \mathbf{2}, -\frac{1}{2} \right), \hspace{2mm}\left( \mathbf{3}, \mathbf{2}, \frac{1}{6} \right),\hspace{2mm} \left( \mathbf{3}, \mathbf{2}, -\frac{5}{6} \right),
\end{equation}
\noindent and their conjugates.  Again, for a discussion on real representations, please see~\citep{BM}, \citep{BBlog}.  

It should be noted that the first two representation spaces align with the transformation properties of the familiar Standard Model Higgs.  They are also closely related to Hughes' quaternionic Higgs, first introduced in~\citep{FH2021(I)}, \citep{FH2021(II)}.

We label these four irreducible representations~(\ref{scalars}) as $h,$ $H,$ $V_q,$ and $V_r,$ respectively.  Accordingly, $\mathbb{H}_V\otimes\mathbb{O}_V$ may be labeled as 
\begin{equation}\begin{array}{lll} \label{VV}
\mathbb{H}_V\otimes\mathbb{O}_V &=&  h^{\uparrow}\hspace{.7mm} {\ell}\uu + h^{\uparrow *} \hspace{.7mm}{\ell}^*\dd \vspace{2mm}\\
&+& h^{\downarrow}\hspace{.7mm} {\ell}\ud - h^{\downarrow*} \hspace{.7mm}{\ell}^*\du \vspace{2mm}\\
&+& H^{\uparrow} \hspace{.7mm} {\ell}\du - H^{\uparrow *} \hspace{.7mm}{\ell}^*\ud \vspace{2mm}\\
&+&  H^{\downarrow}\hspace{.7mm} {\ell}\dd + H^{\downarrow *} \hspace{.7mm}{\ell}^*\uu \vspace{2mm}\\
&+&  V_q^{i\uparrow}\hspace{.7mm} {\alpha_i}\uu - V_q^{i\uparrow *} \hspace{.7mm}\alpha_i^{\dagger}\dd \vspace{2mm}\\
&+&  V_q^{i\downarrow}\hspace{.7mm} {\alpha_i}\ud + V_q^{i\downarrow *} \hspace{.7mm}\alpha_i^{\dagger}\du \vspace{2mm}\\
&+&  V_r^{i\uparrow}\hspace{.7mm} {\alpha_i}\du + V_r^{i\uparrow *} \hspace{.7mm}\alpha_i^{\dagger}\ud \vspace{2mm}\\
&+&  V_r^{i\downarrow}\hspace{.7mm} {\alpha_i}\dd - V_r^{i\downarrow *} \hspace{.7mm}\alpha_i^{\dagger}\uu,\vspace{2mm}\\
\end{array}\end{equation}
\noindent for $h^{\uparrow}, h^{\downarrow}, H^{\uparrow}, H^{\downarrow}, \dots V^{i\downarrow}_r \in\C$.  As with $\mathbb{H}_+\otimes\mathbb{O}_+$ and $\mathbb{H}_-\otimes\mathbb{O}_-$, it may be confirmed that all factors of the complex $i$ drop out.

On one hand, it is of interest to find the familiar Standard Model Higgs representation within $\mathbb{H}_V\otimes\mathbb{O}_V $.  On the other hand, $\mathbb{H}_V\otimes\mathbb{O}_V $ has not  decomposed into the set of irreducible representations corresponding to a third generation.  So has the model then failed?  Or could there be a method with which one may extract a third generation from $\mathbb{H}_V\otimes\mathbb{O}_V$?

\subsection{Cartan Factorization Diagrams}


Consider the \it Cartan vector factorization \rm introduced in equation~(\ref{factors}),
\begin{equation} V = \Psi_+ \Psi_-^{\dagger}.
\end{equation}
\noindent Cartan's vector factorization allows us to make use of division algebraic multiplication in order to recast certain vector representations as the product of certain spinor representations.  It is in this way that we will identify a third generation within $\mathbb{H}_V\otimes\mathbb{O}_V.$ 

As a concrete example, it may be confirmed that there exist $h^{\uparrow}, h^{\downarrow}, \mathcal{V}_{\textup{R}}^3,  \mathcal{V}_{\textup{L}}^3,  \mathcal{E}_{\textup{L}}^3 \in \C$ such that 
\begin{equation}\begin{array}{rll} \label{cvf1}
 h^{\uparrow}\hspace{.7mm} {\ell}\uu \hspace{.5mm}+ \hspace{.5mm}* &=& \left(\mathcal{V}^3_{\textup{R}}\hspace{.7mm} {\ell}\uu \hspace{.5mm}+ \hspace{.5mm}* \right)\left(  \mathcal{V}^3_{\textup{L}}\hspace{.7mm} {\ell}\uu \hspace{.5mm}+\hspace{.5mm} *  \right)^{\dagger},  \vspace{2mm}\\
h^{\downarrow}\hspace{.7mm} {\ell}\ud\hspace{.5mm} +\hspace{.5mm}*   &=& \left(\mathcal{V}^3_{\textup{R}}\hspace{.7mm} {\ell}\uu \hspace{.5mm}+ \hspace{.5mm}* \right)\left(  \mathcal{E}^3_{\textup{L}}\hspace{.7mm} {\ell}\du\hspace{.5mm} + \hspace{.5mm}*  \right)^{\dagger}.
\end{array}\end{equation}
\noindent Here we are using shorthand notation where ``$+\hspace{1mm}*$" means to add the complex conjugate.  Furthermore, $\dagger$ is meant to symbolize the simultaneous application of both the quaternionic and octonionic anti-involutions described in Subsection~(\ref{RCHOintro}).  Equations~(\ref{cvf1}) correspond to the following \it Cartan factorization diagram: \rm 

\begin{figure}[h!]
\begin{center}
\includegraphics[width=5.5cm]{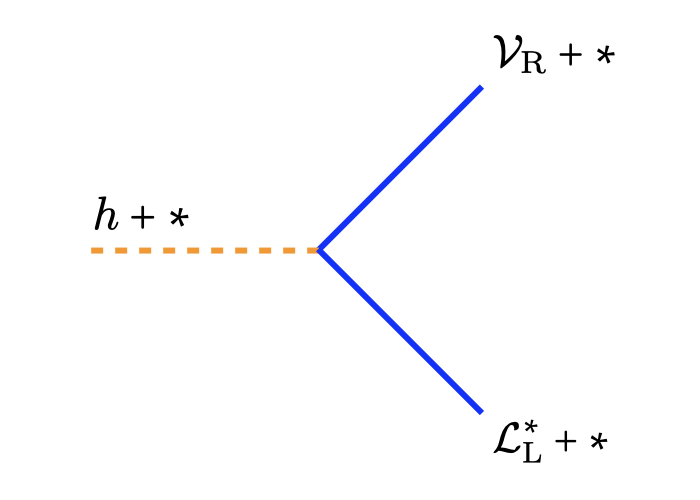}
\caption{\label{CFD1}
A Higgs-to-lepton Cartan factorization diagram.}
\end{center}
\end{figure}
\noindent where $\mathcal{L}_{\textup{L}}$ represents the $\mathfrak{su}(2)_{\textup{L}}$-active lepton doublet.

Similarly, it is possible to find $h^{\uparrow}, h^{\downarrow}, \mathcal{U}_{\textup{R}}^{i3},  \mathcal{U}_{\textup{L}}^{i3},  \mathcal{D}_{\textup{L}}^{i3} \in \C$ such that 
\begin{equation}\begin{array}{rll} \label{cvf2}
 h^{\uparrow}\hspace{.7mm} {\ell}\uu \hspace{.5mm}+\hspace{.5mm} * &=& \left(\mathcal{U}^{i3}_{\textup{R}}\hspace{.7mm} {\alpha_i}\uu \hspace{.5mm}+ \hspace{.5mm}* \right)\left(  \mathcal{U}^{i3}_{\textup{L}}\hspace{.7mm} {\alpha_i}\uu \hspace{.5mm}+\hspace{.5mm} *  \right)^{\dagger},  \vspace{2mm}\\
h^{\downarrow}\hspace{.7mm} {\ell}\ud\hspace{.5mm} +\hspace{.5mm}*   &=& \left(\mathcal{U}^{i3}_{\textup{R}}\hspace{.7mm} {\alpha_i}\uu \hspace{.5mm}+\hspace{.5mm} * \right)\left(  \mathcal{D}^{3i}_{\textup{L}}\hspace{.7mm} {\alpha_i}\du\hspace{.5mm} + \hspace{.5mm}*  \right)^{\dagger},
\end{array}\end{equation}
\noindent where there is no implied sum on the indices intended here.  Equations~(\ref{cvf2}) correspond to the following Cartan factorization diagram:

\begin{figure}[h!]
\begin{center}
\includegraphics[width=5.5cm]{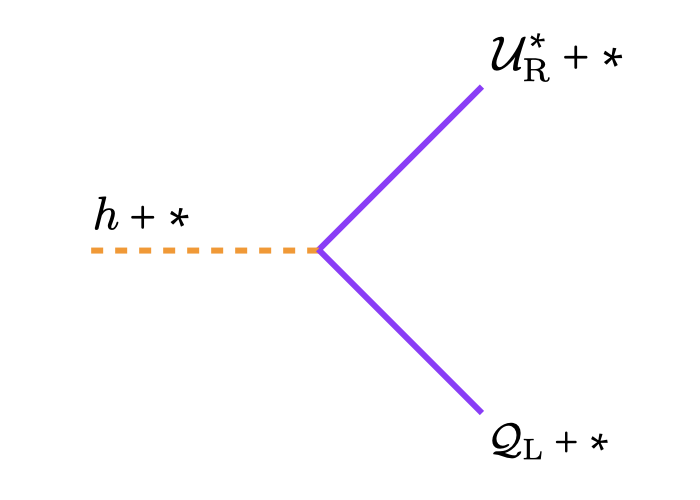}
\caption{\label{CFD2}
A Higgs-to-quark Cartan factorization diagram.}
\end{center}
\end{figure}
\noindent where $\mathcal{Q}_{\textup{L}}$ represents the $\mathfrak{su}(2)_{\textup{L}}$-active quark doublet.

\begin{figure}[h!]
\begin{center}
\includegraphics[width=8cm]{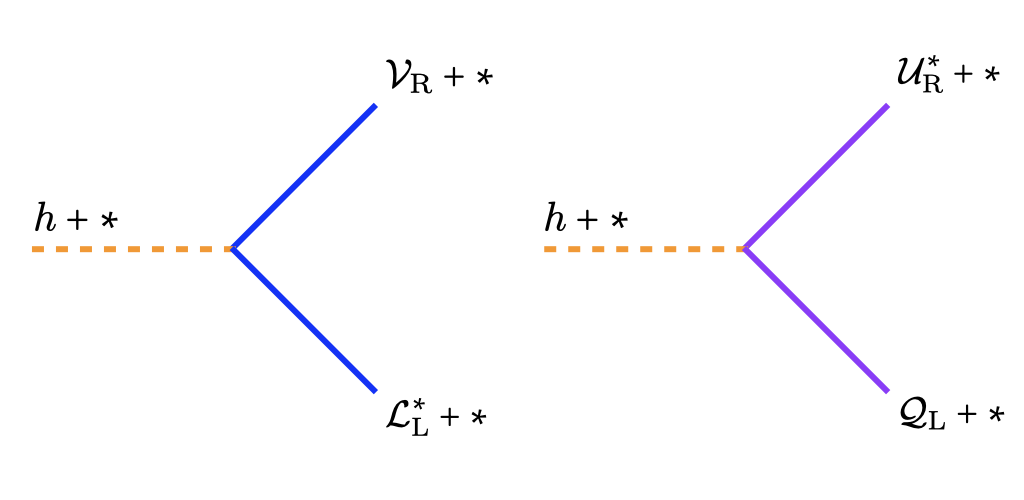}
\includegraphics[width=8cm]{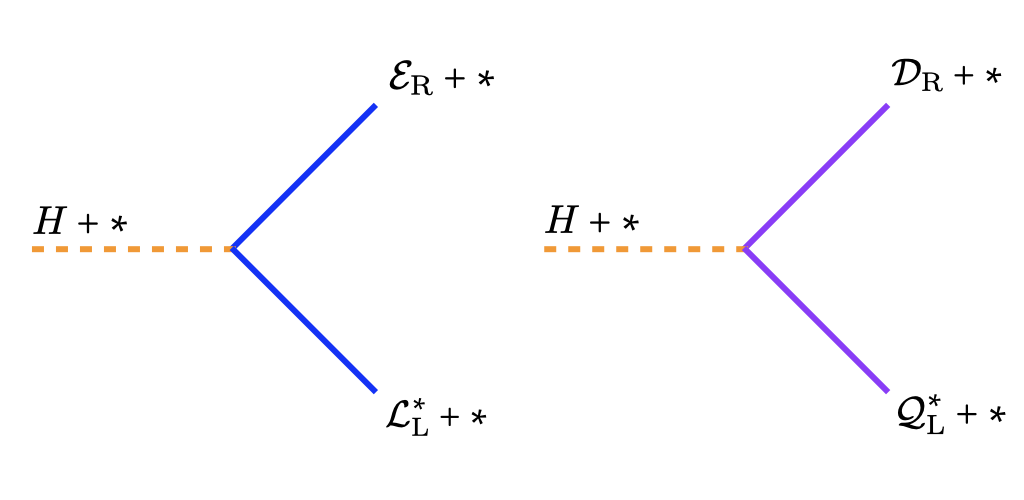}
\caption{\label{HCFD}
The full set of Cartan vector factorization diagrams for $h$ and $H$.  }
\end{center}
\end{figure}

\begin{figure}[h!]
\begin{center}
\includegraphics[width=8cm]{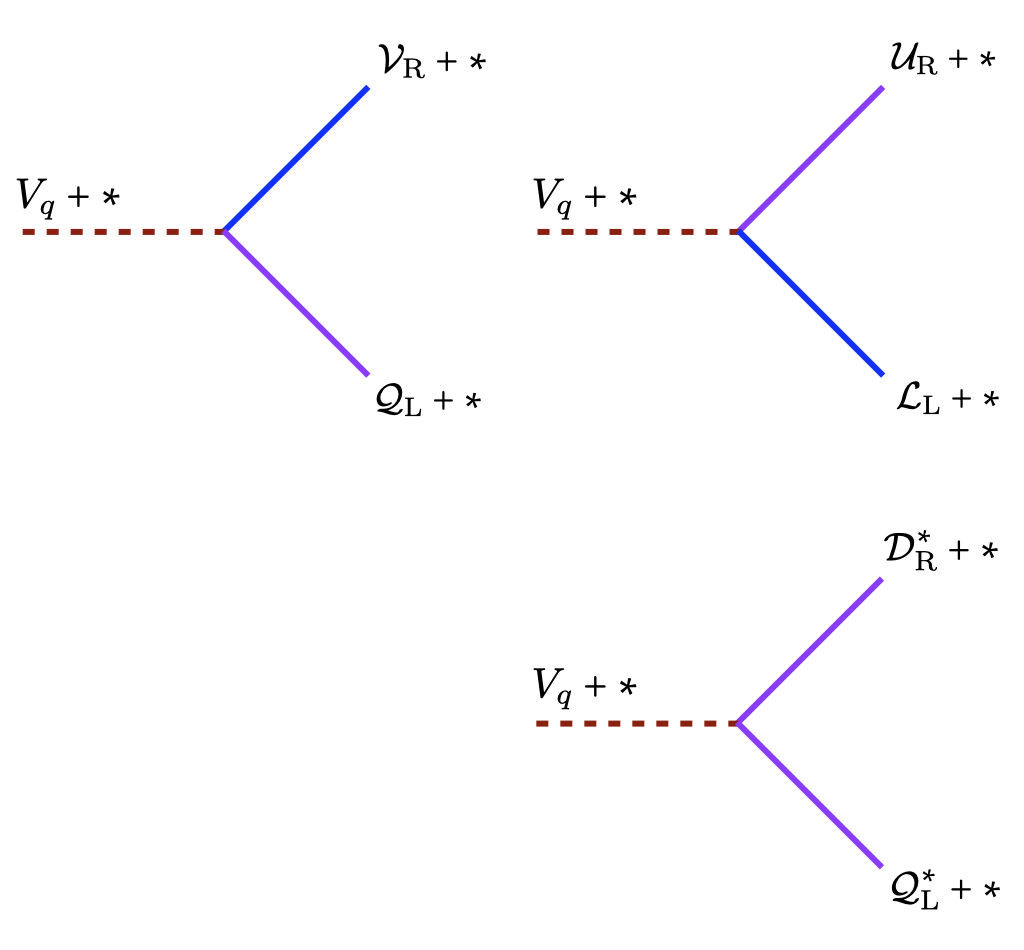}
\caption{\label{VqCFD}
The full set of Cartan vector factorization diagrams for $V_q$.  }
\end{center}
\end{figure}

\begin{figure}[h!]
\begin{center}
\includegraphics[width=8cm]{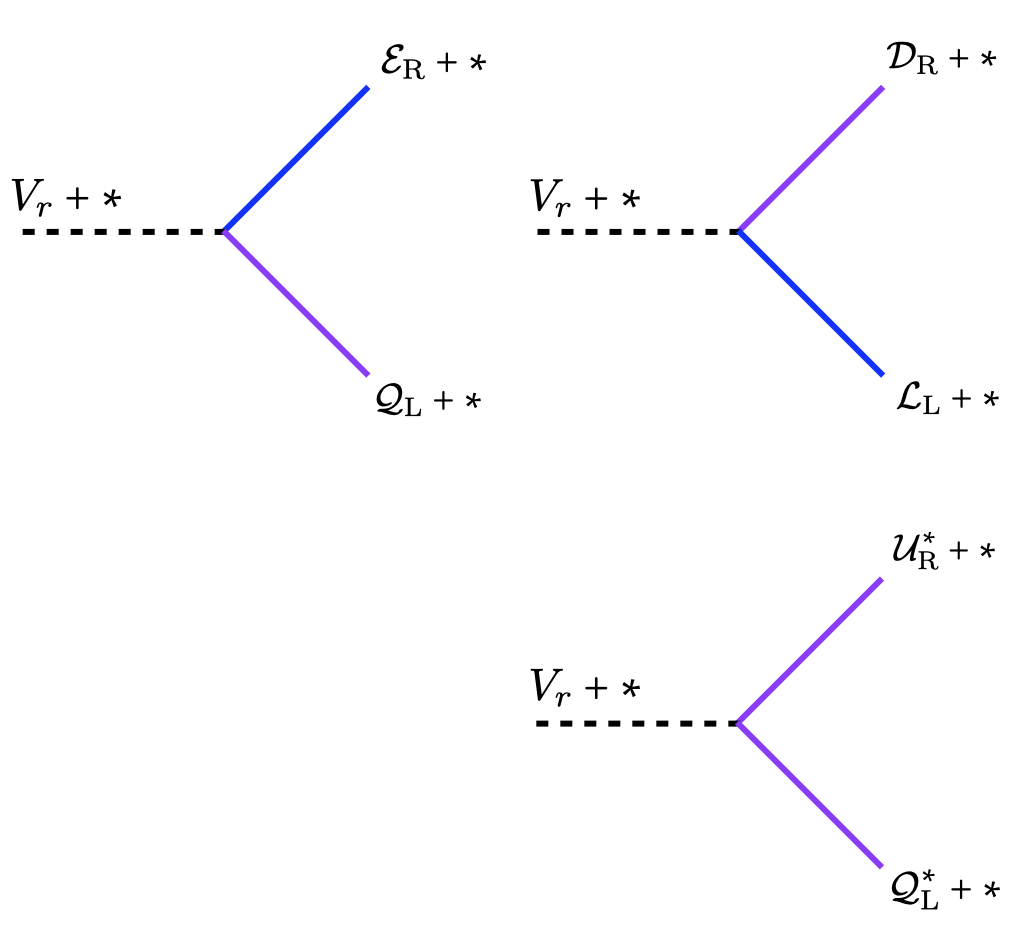}
\caption{\label{VrCFD}
The full set of Cartan vector factorization diagrams for $V_r$.  }
\end{center}
\end{figure}

Energetic readers may wish to calculate all possible  factorizations.  These are too numerous to display here, however, in Figures~\ref{HCFD}, \ref{VqCFD}, and \ref{VrCFD},  we provide  the full set of Cartan vector factorization diagrams.

It is straightforward, although tedious, to show that the $32\hspace{.5mm}\R$ dimensional set of third generation states
\begin{equation}\begin{array}{rll} \label{3rd}
\Psi_+^3 &=& \mathcal{V}^3_{\textup{R}}\hspace{.7mm} {\ell}\uu \hspace{.5mm}+ \hspace{.5mm}* \vspace{2mm}\\
&+& \mathcal{E}^3_{\textup{R}}\hspace{.7mm} {\ell}\du \hspace{.5mm}+ \hspace{.5mm}* \vspace{2mm}\\
&+& \mathcal{U}^{i3}_{\textup{R}}\hspace{.7mm} {\alpha_i}\uu \hspace{.5mm}+ \hspace{.5mm}* \vspace{2mm}\\
&+& \mathcal{D}^{i3}_{\textup{R}}\hspace{.7mm} {\alpha_i}\du \hspace{.5mm}+ \hspace{.5mm}* \vspace{4mm}\\

\Psi_-^3 &=& \mathcal{V}^3_{\textup{L}}\hspace{.7mm} {\ell}\left(\uu+\ud\right) \hspace{.5mm}+ \hspace{.5mm}* \vspace{2mm}\\
&+& \mathcal{E}^3_{\textup{L}}\hspace{.7mm} {\ell}\left(\du+\dd\right) \hspace{.5mm}+ \hspace{.5mm}* \vspace{2mm}\\
&+& \mathcal{U}^{i3}_{\textup{L}}\hspace{.7mm} {\alpha_i}\left(\uu+\ud\right) \hspace{.5mm}+ \hspace{.5mm}* \vspace{2mm}\\
&+& \mathcal{D}^{i3}_{\textup{L}}\hspace{.7mm} {\alpha_i}\left(\du+\dd\right) \hspace{.5mm}+ \hspace{.5mm}* 
\end{array}\end{equation}
\noindent produces the full set of Cartan factorization diagrams in Figures~\ref{HCFD}, \ref{VqCFD}, \ref{VrCFD}.







\subsection{Complexification \label{comp}}

In order to describe unconstrained off-shell degrees of freedom (or alternatively the complex Hilbert space described in~\citep{AGUTS})  we may complexify these three generations so as to obtain
$$(\C_+\otimes\mathbb{H}_+\otimes\mathbb{O}_+)\oplus (\C_-\otimes\mathbb{H}_-\otimes\mathbb{O}_-) \oplus (\C_V\otimes\mathbb{H}_V\otimes\mathbb{O}_V).$$ 

In this case, equations~(\ref{++}), (\ref{--}), and (\ref{VV}) become
\begin{equation}\begin{array}{lll} 
\C_+\otimes\mathbb{H}_+\otimes\mathbb{O}_+ &=&  \mathcal{V}^1_{\textup{R}}\hspace{.7mm} {\ell}\uu + \bar{\mathcal{V}}^{1}_{\textup{R}} \hspace{.7mm}{\ell}^*\dd \vspace{2mm}\\
&+&  \mathcal{V}^2_{\textup{R}}\hspace{.7mm} {\ell}\ud - \bar{\mathcal{V}}^{2}_{\textup{R}} \hspace{.7mm}{\ell}^*\du  \vspace{2mm}\\
&+& \dots,
\end{array}\end{equation}

\begin{equation}\begin{array}{lll} 
\C_-\otimes\mathbb{H}_-\otimes\mathbb{O}_- &=&  \mathcal{V}^1_{\textup{L}}\hspace{.7mm} {\ell}\uu + \bar{\mathcal{V}}^{1}_{\textup{L}} \hspace{.7mm}{\ell}^*\dd \vspace{2mm}\\
&+&  \mathcal{V}^2_{\textup{L}}\hspace{.7mm} {\ell}\ud - \bar{\mathcal{V}}^{2}_{\textup{L}} \hspace{.7mm}{\ell}^*\du \vspace{2mm}\\
&+& \dots,

\end{array}\end{equation}

\begin{equation}\begin{array}{lll} 
\C_V\otimes\mathbb{H}_V\otimes\mathbb{O}_V &=&  h^{\uparrow}\hspace{.7mm} {\ell}\uu + \bar{h}^{\uparrow } \hspace{.7mm}{\ell}^*\dd \vspace{2mm}\\&+& h^{\downarrow}\hspace{.7mm} {\ell}\ud - \bar{h}^{\downarrow} \hspace{.7mm}{\ell}^*\du \vspace{2mm}\\
&+& \dots,\vspace{2mm}
\end{array}\end{equation}
\noindent where the complex coefficients written with bars overtop are now independent from the unbarred complex coefficients.  It is worth noting that the Cartan factorizations of the previous subsection carry over in an obvious manner to the complexified case.

Upon complexification, it is natural to consider $\mathfrak{tri}(\C)\oplus\mathfrak{tri}(\mathbb{H})\oplus\mathfrak{tri}(\mathbb{O}),$ as first proposed in~\citep{FH2021(II)}.  Pushing the idea one step further, in future work we will explore a possible extension of the triality construction of the Freudenthal-Tits magic square.  The  vector space associated with this new Lie algebra is written as
\begin{equation}\begin{array}{c}\mathfrak{tri}(\C)\oplus\mathfrak{tri}(\mathbb{H})\oplus\mathfrak{tri}(\mathbb{O}) \vspace{2mm}\\ \oplus \vspace{2mm}\\
(\C_+\otimes\mathbb{H}_+\otimes\mathbb{O}_+)\oplus (\C_-\otimes\mathbb{H}_-\otimes\mathbb{O}_-) \oplus (\C_V\otimes\mathbb{H}_V\otimes\mathbb{O}_V),\vspace{2mm} \end{array}\end{equation}
\noindent whose action generalizes that of $ \mathfrak{e}_7 \simeq \left(\mathfrak{tri}(\mathbb{H})\oplus\mathfrak{tri}(\mathbb{O})\right) + 3\left(\HO\right)$  in the obvious way, \citep{mia}.  Alternatively, one may consider $ \mathfrak{e}_7$'s complexification.  

\it Again, we mention that Hodge duality and a requirement of anomaly cancellation may be used to reduce symmetries. \rm



\subsection{Yukawa terms and multiple Higgs}

It is straightforward to confirm that the representations of Subsection~\ref{comp} may be assembled into Yukawa scalars of the following types:
\begin{equation}\begin{array}{cc}   \label{yuk}
\langle \mathcal{V}^{\dagger}_{\textup{R}}\hspace{.5mm} {h} \hspace{.2mm}{\mathcal{L}}_{\textup{L}} \rangle, & \langle \mathcal{E}^{\dagger}_{\textup{R}}\hspace{.5mm} {H} \hspace{.5mm}{\mathcal{L}}_{\textup{L}} \rangle,  \vspace{2mm} \\

\langle \mathcal{U}^{\dagger}_{\textup{R}}\hspace{.5mm} {h} \hspace{.2mm}{\mathcal{Q}}_{\textup{L}} \rangle, & \langle \mathcal{D}^{\dagger}_{\textup{R}}\hspace{.5mm} {H} \hspace{.5mm}{\mathcal{Q}}_{\textup{L}} \rangle,  \vspace{2mm} \\

\langle \mathcal{V}^{\dagger}_{\textup{R}}\hspace{.5mm} {V_q^*} \hspace{.2mm}{\mathcal{Q}}_{\textup{L}} \rangle, & \langle \mathcal{E}^{\dagger}_{\textup{R}}\hspace{.5mm} {V_r} \hspace{.5mm}{\mathcal{Q}}_{\textup{L}}^* \rangle,  \vspace{2mm} \\

\langle \mathcal{U}^{\dagger}_{\textup{R}}\hspace{.5mm} {V_q} \hspace{.2mm}{\mathcal{L}}_{\textup{L}}^* \rangle, & \langle \mathcal{D}^{\dagger}_{\textup{R}}\hspace{.5mm} {V_r} \hspace{.5mm}{\mathcal{L}}_{\textup{L}}^* \rangle,  \vspace{2mm} \\

\langle \mathcal{D}^{\dagger}_{\textup{R}}\hspace{.5mm} {V_q^*} \hspace{.2mm}{\mathcal{Q}}_{\textup{L}}^* \rangle, & \langle \mathcal{U}^{\dagger}_{\textup{R}}\hspace{.5mm} V_r^* \hspace{.5mm}{\mathcal{Q}}_{\textup{L}}^* \rangle,  \vspace{2mm} \\

\end{array}\end{equation}
\noindent and their conjugates. Here $ \mathcal{V}_{\textup{R}}$ transforms as $\left(\mathbf{1}, \mathbf{1}, 0\right)$, $h$  transforms as $\left( \mathbf{1}, \mathbf{2}, \frac{1}{2}\right),$ $\mathcal{L}_{\textup{L}}$ transforms as $\left(\mathbf{1}, \mathbf{2}, -\frac{1}{2}\right),$ etc.  

A few comments are in order.  First of all, we point out that in the complexified case, there are effectively four $\left( \mathbf{1}, \mathbf{2}, -\frac{1}{2} \right)$ Higgs representations, $h^*,$ $H,$ $\bar{h},$ $\bar{H}^*,$ in contrast to the Standard Model's single Higgs.  (Incidentally, there are four masses per generation.)  

Secondly, notice that the preservation of $h^*$ and $H$ as separate representations breaks $\mathfrak{su}(2)_{\textup{R}}$ symmetry, thereby ensuring \it chirality.  \rm

Finally, we point out that while the $4\hspace{.5mm}\R$ dimensional $h^*$ and $H$ bosons align relatively closely with the Standard Model's phenomenology, the $12\hspace{.5mm}\R$ dimensional   $V_q$  and $V_r$  bosons do not.  Could there be a constraint that suppresses or eliminates the effect of $V_q$ and $V_r$ while preserving the effect of the Higgs representations?


\subsection{The Non-Degenerate Trilinear Form (NDTF) Constraint}

Suppose we require that all Yukawa terms originate from a \it non-degenerate \rm trilinear form 
\begin{equation}t(\Psi_+, V, \Psi_- ) = \langle \Psi_+^{\dagger}\hspace{.5mm} V\hspace{.5mm} \Psi_- \rangle
\end{equation}
\noindent of Section~\ref{trilin}.  Recall that this constraint requires that $\Psi_+, V, \Psi_- $ each live in algebras isomorphic to $\R,$ $\C,$ $\mathbb{H},$ or $\mathbb{O}$.  

Then it is straightforward to see that certain linear combinations of $h,$ $h^*,$ $\bar{h},$ $\bar{h}^*,$ $H,$  $H^*,$ $\bar{H},$ and $\bar{H}^*$ accommodate quaternionic versions of this trilinear form.  For a detailed example of one such quaternionic triality Higgs system, please see~\citep{FH2021(II)}.  

In contrast, with their $12\hspace{.5mm}\R$ dimensions, no linear combination of  $V_q,$ $V_q^*,$ $\bar{V}_q,$ $\bar{V}_q^*,$ $V_r,$ $V_r^*,$ $\bar{V}_r,$ $\bar{V}_r^*$  acts on fermions as $\R$, $\C$, $\mathbb{H}$, or $\mathbb{O}$.  Hence, we would expect these Yukawa terms to be suppressed or excluded at energy levels where the Standard Model's $\gsm$ symmetries are relevant.  It is of interest to note that the $\left( \mathbf{1}, \mathbf{2}, \pm\frac{1}{2}\right),$ Higgs representations constitute the minimal set within $\C_V\otimes\mathbb{H}_V\otimes\mathbb{O}_V$ necessary to produce all of the third-generation fermions states.

In continuing work, we explore how this \emph{non-degenerate trilinear form (NDTF) constraint} may be used to build a Lagrangian in the context of a field theory.

\section{Summary}

In this article, we started out by identifying the Standard Model's $\mathfrak{su}(3)_{\textup{C}}\oplus\mathfrak{su}(2)_{\textup{L}}\oplus\mathfrak{u}(1)_{\textup{Y}}$ symmetries within the triality symmetries $\mathfrak{tri}(\mathbb{H})\oplus \mathfrak{tri}(\mathbb{O})\subset \mathfrak{tri}(\mathbb{C})\oplus\mathfrak{tri}(\mathbb{H})\oplus \mathfrak{tri}(\mathbb{O}).$  We applied these Standard Model symmetries onto $(\mathbb{H}_+\otimes\mathbb{O}_+)\oplus (\mathbb{H}_-\otimes\mathbb{O}_-) \oplus (\mathbb{H}_V\otimes\mathbb{O}_V)$, and subsequently its complexification $(\C_+\otimes\mathbb{H}_+\otimes\mathbb{O}_+)\oplus (\C_-\otimes\mathbb{H}_-\otimes\mathbb{O}_-) \oplus (\C_V\otimes\mathbb{H}_V\otimes\mathbb{O}_V)$.  We found that together, the spinor and conjugate spinor representations gave two generations of Standard Model irreps, including sterile neutrinos.  However, the vector representations did not immediately produce the irreps of a third generation.  Instead, one finds 
\begin{equation}\begin{array}{cccc}
\label{setscalars}
\left( \mathbf{1}, \mathbf{2}, \frac{1}{2} \right), &\left( \mathbf{1}, \mathbf{2}, -\frac{1}{2} \right), &\left( \mathbf{3}, \mathbf{2}, \frac{1}{6} \right),& \left( \mathbf{3}, \mathbf{2}, -\frac{5}{6} \right),\vspace{2mm}\\
\left( \mathbf{1}, \mathbf{2}, -\frac{1}{2} \right), &\left( \mathbf{1}, \mathbf{2}, \frac{1}{2} \right), &\left( \mathbf{3}^*, \mathbf{2}, -\frac{1}{6} \right),& \left( \mathbf{3}^*, \mathbf{2}, \frac{5}{6} \right)\vspace{2mm}
\end{array}\end{equation}
\noindent in the complexified case. Notably, this set includes four copies of the familiar Standard Model Higgs representations.  

It turns out that a third generation may be identified within $\C_V\otimes\mathbb{H}_V\otimes\mathbb{O}_V$ by making use of \it Cartan factorization, \rm which allows vector representations to be factorized into spinor and conjugate spinor representations.  We demonstrate that this method yields a third generation.

Within the set of $\C_V\otimes\mathbb{H}_V\otimes\mathbb{O}_V$ irreps, the Standard Model Higgs are found to constitute the smallest  representations necessary to produce a full generation of fermions.  With this said, our model, at least na\"{i}vely, seems to make a significant departure from the Standard Model.  Instead of one Higgs with four Yukawa coupling constants across a generation, one finds four  Higgs representations.  

Future work will explore the other Cartan factorizations, $\Psi_+ = V\hspace{.5mm} \Psi_-$ and $\Psi_-^{\dagger} =\Psi_+^{\dagger}\hspace{.5mm}V,$ in the context of the first two generations.

Finally, we point out that Yukawa terms involving the above $\left( \mathbf{3}, \mathbf{2}, \frac{1}{6} \right)$ and $\left( \mathbf{3}, \mathbf{2}, -\frac{5}{6} \right)$ representations may be excluded, while Yukawa terms involving $\left( \mathbf{1}, \mathbf{2}, \pm\frac{1}{2} \right)$  may be preserved, if one requires that our Lagrangian's Yukawa terms are built only from non-degenerate trilinear forms.  This requirement is introduced as the \emph{non-degenerate trilinear form (NDTF) constraint}.  Dynamics and symmetry breaking in this model are subjects of current investigation.










\begin{acknowledgements} \it To J\'{o}zef's first three steps. \rm  The authors are grateful for conversations with Michele Galli, Rob Klabbers, Jens K\"{o}plinger, Andy Randono,  and Carlos Tamarit.
This work is graciously supported by the VW Stiftung Freigeist Fellowship, Humboldt-Universit\"{a}t zu Berlin, and the University of Leeds. 

\end{acknowledgements}


\medskip

\end{document}